%% file: jas.tex
\documentclass[10pt,twocolumn,a4paper]{article}

\usepackage{epsfig}
\usepackage{graphicx}
\usepackage{amssymb}
\usepackage{amsmath}

\newcommand{\code}[1]{\texttt{#1}}
\begin{document}

\title{Parallel and distributed Gr\"obner bases computation in JAS}

\author{Heinz Kredel\\
{\small IT-Center, University of Mannheim}\\
\code{kredel@rz.uni-mannheim.de}
}

\def\today{18. July 2010}

\maketitle
\thispagestyle{empty}

\begin{abstract} 
  This paper considers parallel Gr\"obner bases algorithms on
  distributed memory parallel computers with multi-core compute nodes.
  We summarize three different Gr\"obner bases implementations: shared
  memory parallel, pure distributed memory parallel and distributed
  memory combined with shared memory parallelism.  The last algorithm,
  called distributed hybrid, uses only one control communication
  channel between the master node and the worker nodes and keeps
  polynomials in shared memory on a node. The polynomials are
  transported asynchronous to the control-flow of the algorithm in a
  separate distributed data structure.  The implementation is generic
  and works for all implemented (exact) fields.  We present new
  performance measurements and discuss the performance of the
  algorithms.
\end{abstract}




\section{Introduction} 


We summarize parallel algorithms for computing Gr\"obner bases on
todays cluster parallel computing environments in a Java computer
algebra system (JAS), which we have developed in the last years
\cite{Kredel:2009,Kredel:2010}.  Our target hardware are distributed
memory parallel computers with multi-core compute nodes.  Such
computing infrastructure is predominant in todays high performance
computing clusters (HPC).  The implementation of Gr\"obner bases
algorithms is part of the essential building blocks for any
computation in algebraic geometry.  Our aim is an implementation in a
modern object oriented programming language with generic data types,
as it is provided by Java programming language.

Besides the sequential algorithm, we consider three 
Gr\"obner bases implementations: multiple threads using shared memory,
pure distributed memory with communication of polynomials between
compute nodes and distributed memory combined with multiple threads on
the nodes. The last algorithm, called distributed hybrid, uses only
one control communication channel between the master node and the
worker nodes and keeps polynomials in shared memory on a node.  The
polynomials are transported asynchronous to the control-flow of the
algorithm in a separate distributed data structure.  In this paper we
present new performance measurements on a grid-cluster
\cite{bwgrid:2008} and discuss performance of the algorithms.

An object oriented design of a Java computer algebra system (called
JAS) as type safe and thread safe approach to computer algebra is
presented in \cite{Kredel:2000,Kredel:2006,Kredel:2007,Kredel:2008b}.
JAS provides a well designed software library using generic types for
algebraic computations implemented in the Java programming language.
The library can be used as any other Java software package or it can
be used interactively or interpreted through an Jython (Java Python)
front-end. The focus of JAS is at the moment on commutative and
solvable polynomials, Gr\"obner bases, greatest common divisors and
applications.  JAS contains interfaces and classes for basic
arithmetic of integers, rational numbers and multivariate polynomials
with integer or rational number coefficients.

\subsection{Parallel Gr\"obner bases} 

The computation of Gr\"obner bases (via the Buchberger algorithm)
solves an important problem for computer algebra \cite{Becker:1993}.
These bases play the same role for the solution of systems of
algebraic equations as the LU-de\-compo\-sit\-ion, obtained by
Gaussian elimination, for systems of linear equations.  Unfortunately
the computation of such polynomial bases is notoriously hard, both
with sequential and parallel algorithms.  So any improvement of this
algorithm is of great importance.  For a discussion of the problems
with parallel versions of this algorithm, see the introduction in
\cite{Kredel:2009}.

\subsection{Related work} 
\label{sec:rela}

In this section, we briefly summarize the related work.  Related work
on computer algebra libraries and an evaluation of the JAS library in
comparison to other systems can be found in
\cite{Kredel:2007,Kredel:2008a}. 

Theoretical studies on parallel computer algebra focus on parallel
factoring and problems which can exploit parallel linear
algebra 
\cite{Lenstra:1992,zurGathen:1984}.
Most reports on experiences and results of parallel computer algebra
are from systems written from scratch or where the system source code
was available.  A newer approach of a multi-threaded polynomial
library implemented in C is for example \cite{ElDinTrebuchet:2007}.
From the commercial systems some reports are about
Maple \cite{Char:1990} 
(workstation clusters), and Reduce
\cite{Neun:1989} 
(automatic compilation, vector processors).
Multi-processing support for Aldor (the programming language of Axiom) 
is presented in \cite{MorenoMazaWatt:2007}.
Grid aware computer algebra systems are for example
\cite{GridMath:2008}.  The SCIEnce project works on Grid facilities
for the symbolic computation systems GAP, KANT, Maple and MuPAD
\cite{SCIEnce:2009,SymGrid:2007}.  
Java grid middle-ware systems and parallel computing platforms are
presented in
\cite{Ibis:2010,GridGain:2008,Caromel:2008,Kaminski:2010,AtejiPX:2010,mpj:2010}.
For further overviews see section 2.18 in the report
\cite{Grabmaier:2003} and the tutorial \cite{RochVillard:1997}.

For the parallel Buchberger algorithm the idea of parallel reduction
of S-polynomials seems to be originated by Buchberger and was in the
folklore for a while.  First implementations have been reported, for
example, by Hawley \cite{Hawley:1992} and others
\cite{Leykin:2004,Yanovich:2008,InoueSato:2007}.  For triangular
systems multi-threaded parallel approaches have been reported by
\cite{LiMorenoMaza:2007,MorenoMazaXie:2007}.

\subsection{Outline} 

Due to limited space we must assume that you are familiar with Java,
object oriented programming and mathematics of Gr\"obner bases
\cite{Becker:1993}.  Section 2 introduces the expected and developed
infrastructure to implement parallel and distributed Gr\"obner bases.
The Gr\"obner base algorithms are summarized in section 3.  Section 4
evaluates several aspects of the design, namely termination detection,
performance, the `workload paradox', and selection strategies.
Finally section 5 draws some conclusions and shows possible future
work.

For the convenience, this paper contains summaries and revised parts
of \cite{Kredel:2009,Kredel:2010} to explain the new performance
measurements.  Performance figures and tables are presented throughout
the paper.  Explanations are in section \ref{:sec:perf}.

\section{Hard- and middle-ware} 

In this section we summarize computing hardware and middle-ware
components required for the implementation of the presented
algorithms.  The suitability of the Java computing platform for
parallel computer algebra has been discussed for example in
\cite{Kredel:2009,Kredel:2010}.

\subsection{Hardware} 
\label{sec:hardware}

Common grid computing infrastructure consists of nodes of multi-core
CPUs connected by a high-performance network. We have access to the
bwGRiD infrastructure \cite{bwgrid:2008}.
It consists of $2 \times 140$ $8$-core CPU nodes
at $2.83$ GHz with $16$ GB main memory connected by a $10$ Gbit InfiniBand and
$1$ Gbit Ethernet network.  The operating system is Scientific Linux 5.0
and has shared Lustre home directories and a PBS batch system with
Maui scheduler.

The performance of the distributed algorithms depend on the fast
Infini\-Band networking hard\-ware. We have done performance tests
also with normal Ether\-net networking hard\-ware.  The Ethernet
connection of the nodes in the bw\-GRiD cluster is $1$ Gbit to a
switch in a blade center containing $14$ nodes and a $1$ Gbit
connection of the blade centers to a central switch.  We could not
obtain any speedup for the distributed algorithm on an Ether\-net
connection, only with the Infini\-Band connection a speedup can be
reported.

The Infini\-Band connection is used with the TCP/IP protocol. The
support for the direct Infini\-Band protocol, by-passing the TCP/IP
stack, will eventually be available in JDK 1.7 in 2010/11. The evaluation
of the direct Infini\-Band access will be future work.

\subsection{Execution middle-ware} 

In this section we summarize the execution and communication
middle-ware used in our implementation, for details see
\cite{Kredel:2009,Kredel:2010}.  The execution middle-ware is general
purpose and independent of a particular application like Gr\"obner
bases and can thus be used for many kinds of algebraic algorithms.

\begin{figure}[thb]
\centering
\epsfig{file=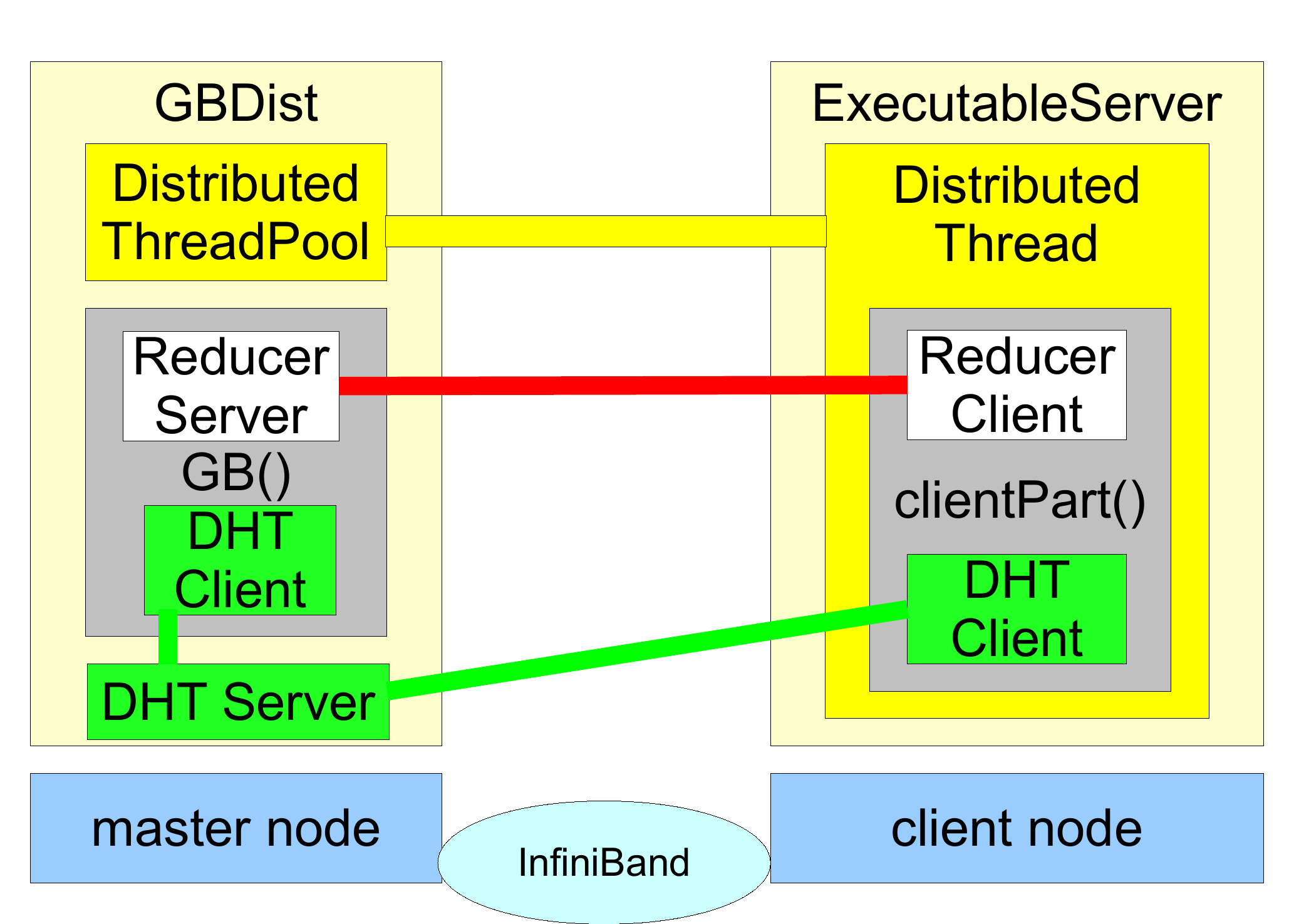,clip=,width=1.0\linewidth}
\caption{Middleware overview.}
\label{fig:middleware}
\end{figure}

The infrastructure for the distributed partner processes uses a daemon
process, which has to be setup via the normal cluster computing tools
or some other means. The cluster tools available at Mannheim use PBS
(portable batch system). PBS maintains a list of nodes allocated for a
cluster job which is used 
in a loop with \code{ssh}-calls to start the daemon on the available
compute nodes.  The lowest level class \code{Exe\-cut\-able\-Server}
implements the daemon processes, see figure \ref{fig:middleware}.
They receive serialized instances of classes which implement the
\code{Remote\-Executable} interface and execute them (call their
\code{run()} method).  On top of the low level daemons is a thread
pool infrastructure, which distributes jobs to the remote daemons, see
classes \code{Dist\-Thread\-Pool} and \code{Dist\-Pool\-Thread} in
figure \ref{fig:middleware}.

The communication infrastructure is provided on top of TCP/IP sockets
with Java object serialization. In case of the distributed hybrid
algorithm we have only one TCP/IP connection (for control) between the
master and the remote threads. To be able to distinguish messages
between specific threads on both sides we use tagged messages
channels.  Each message is send together with a tag (an unique
identifier) and the receiving side can then wait only for messages
with specific tags. For details on the implementation and alternatives
see \cite{Kredel:2010}.

\subsection{Data structure middle-ware} 

We try to reduce communication cost by employing a distributed data
structure with asynchronous communication which can be overlapped with
computation.  Using marshalled objects for transport, the object
serialization overhead is minimized.  This data structure middle-ware
is independent of a particular application like Gr\"obner bases and
can be used for many kinds of applications.

The distributed data structure is implemented by class
\code{Dist\-Hash\-Table}, called `DHT client' in figure
\ref{fig:middleware}, see also figure \ref{fig:threadnode}. It
implements the \code{Map} interface and extends \code{Abs\-tract\-Map}
from the \code{java.util} package with type parameters and can so be
used in a type safe way. In the current program version we use a
centralized control distributed hash table, a decentralized version
will be future work.  For the usage of the data structure the clients
only need the network node name and port of the master.  In addition
there are methods like \code{getWait()}, which expose the different
semantics of a distributed data structure as it blocks until an
element for the key has arrived.

\section{Gr\"obner bases} 

In this section we summarize the sequential, the shared memory
parallel and distributed versions of algorithms to compute Gr\"obner
bases as described in \cite{Kredel:2009,Kredel:2010}.  For the
mathematics of the sequential version of the Buchberger algorithm see
\cite{Becker:1993} or other books.

\begin{figure}[thb]
\centering
\epsfig{file=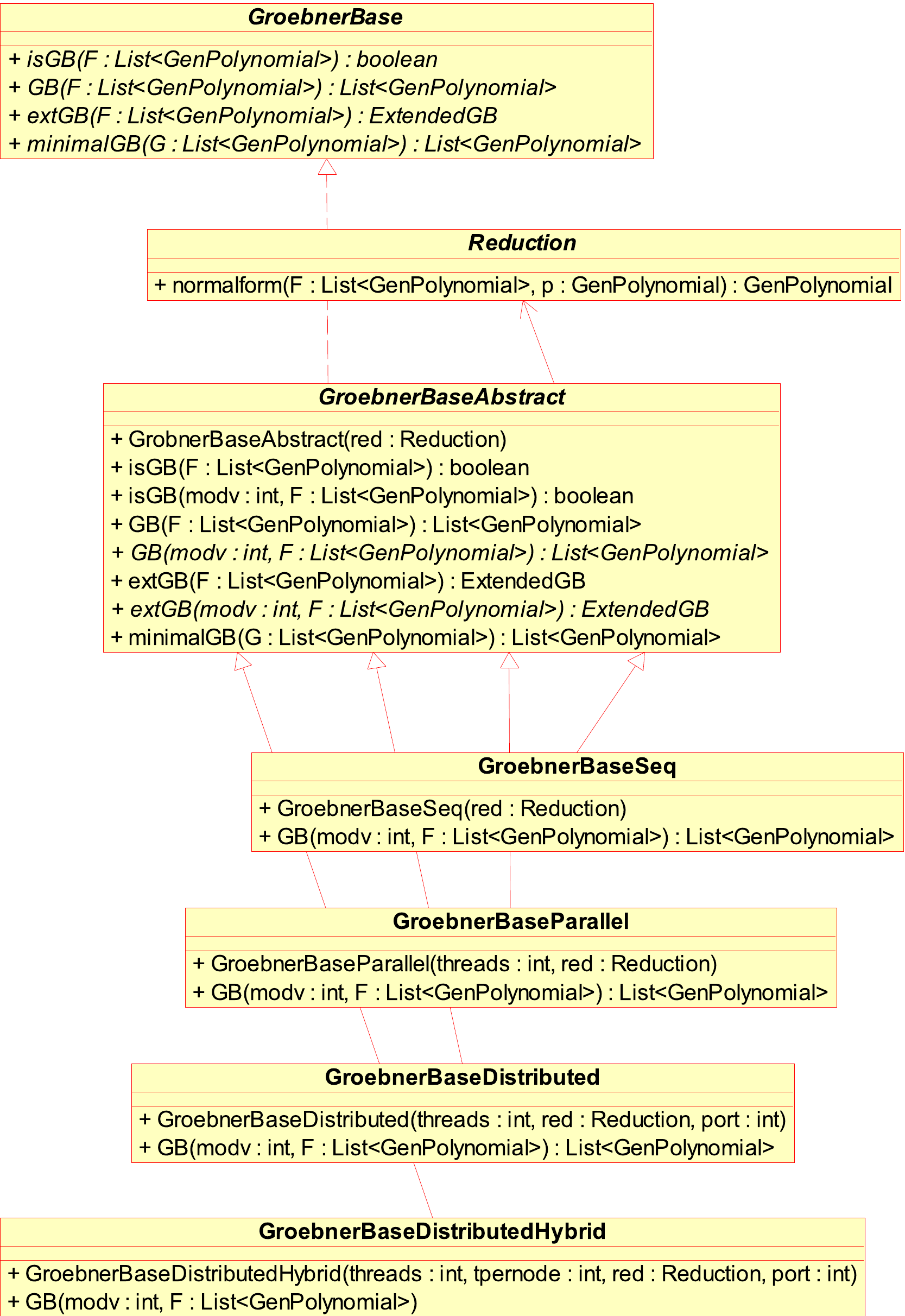,clip=,width=0.950\linewidth}
\caption{UML diagram of Gr\"obner base classes.}
\label{fig:gbase}
\end{figure}

\subsection{Sequential Gr\"obner bases} 

The sequential algorithm takes a set of (multivariate) polynomials
over a field as input and produces a new set of polynomials which
generates the same polynomial ideal but additionally the reduction
relation with respect to the new set of polynomials has unique normal
forms.  The implementation is generic and works for all (exact) fields
implemented in JAS and also handles the case of modules over
polynomial rings.  In the algorithm, first a set of critical pairs is
generated, then the S-polynomial of each critical pair is checked if
it can be reduced to zero. If not, the resulting reduction rest is
added to the set of polynomials and new critical pairs are generated.
The algorithm terminates if all S-polynomials of critical pairs reduce
to zero (which is guarantied to happen by Dickson's lemma).  The
implementation only uses Buchberger's first and second criterion (see
\cite{Becker:1993}). Optimizations like the F4 or F5 algorithm
\cite{GebauerMoeller:1988,GioviniMora:1991,Faugere:1999,Faugere:2002}
are not incorporated. In this paper we focus on the comparison of the
`simple' sequential Buchberger algorithm to `simple' parallel and
distributed algorithms without interference with further
optimizations. Optimized algorithms will be studied and compared in
future work.

The implementation of the parallel and distributed versions is based
on the sequential algorithm.  These algorithms are implemented
following standard object oriented patterns (see figure
\ref{fig:gbase}). There is an interface, called \code{Groebner\-Base},
which specifies the desirable functionality.  Then there is an
abstract class, called \code{Groebner\-Base\-Abs\-tract}, which
implements as many methods as possible.  Finally there are concrete
classes which extend the abstract class and implement different
algorithmic details.  For example \code{Groeb\-ner\-Base\-Seq}
implements a sequential, \code{Groeb\-ner\-Base\-Parallel} implements
a thread parallel, \code{Groeb\-ner\-Base\-Distri\-buted} implements a
network distributed version of the Gr\"obner base algorithm as
described in \cite{Kredel:2009}.
\code{Groeb\-ner\-Base\-Distri\-buted\-Hybrid} implements the hybrid
algorithm as described in \cite{Kredel:2010}.

The polynomial reduction algorithms are implemented by me\-thods
\code{nor\-mal\-form()} in classes \code{Re\-duc\-tion\-Seq} and
\code{Re\-duc\-tion\-Par}. The later class does not implement a parallel
reduction algorithm, as its name may suggest, but a sequential
algorithm which can tolerate and use asynchronous updates of the
polynomial list by other threads.  A parallel reduction implementation
is still planed for future work.

\subsection{Parallel Gr\"obner bases} 

The shared memory parallel Gr\"obner bases algorithm is a variant of
the classical sequential Buchberger algorithm and follows our previous
work in \cite{Kredel:1994}.  It maintains a shared data structure,
called pair list, for book keeping of the computations.  This data
structure is implemented by classes \code{Critical\-Pair\-List} and
\code{Ordered\-Pair\-List}, see section \ref{sec:select}.  Both have
synchronized methods \code{put()} and \code{get\-Next()} respectively
\code{remove\-Next()} to update the data structure and acquire a pair
for reduction. In this way the pair list is used as work queue in the
parallel and the distributed implementations.  As long as there are
idle threads, critical pairs are taken from the work queue and
processed in a thread. The processing consists of forming
S-polynomials and doing polynomial reductions with respect to the
current list of polynomials. When a reduction finished and the result
polynomial is non-zero, new critical pairs are formed and the
polynomial is added to the list of polynomials. Note, due to different
computing times needed for reduction of polynomials the results may
finish in a different sequence order than in the sequential algorithm,
see section \ref{sec:select}. As the proof of the `simple' Buchberger
algorithm does not depend on a certain sequence order, the correctness
of the parallel and distributed algorithms is established.

\begin{figure}[thb]
\centering
\epsfig{file=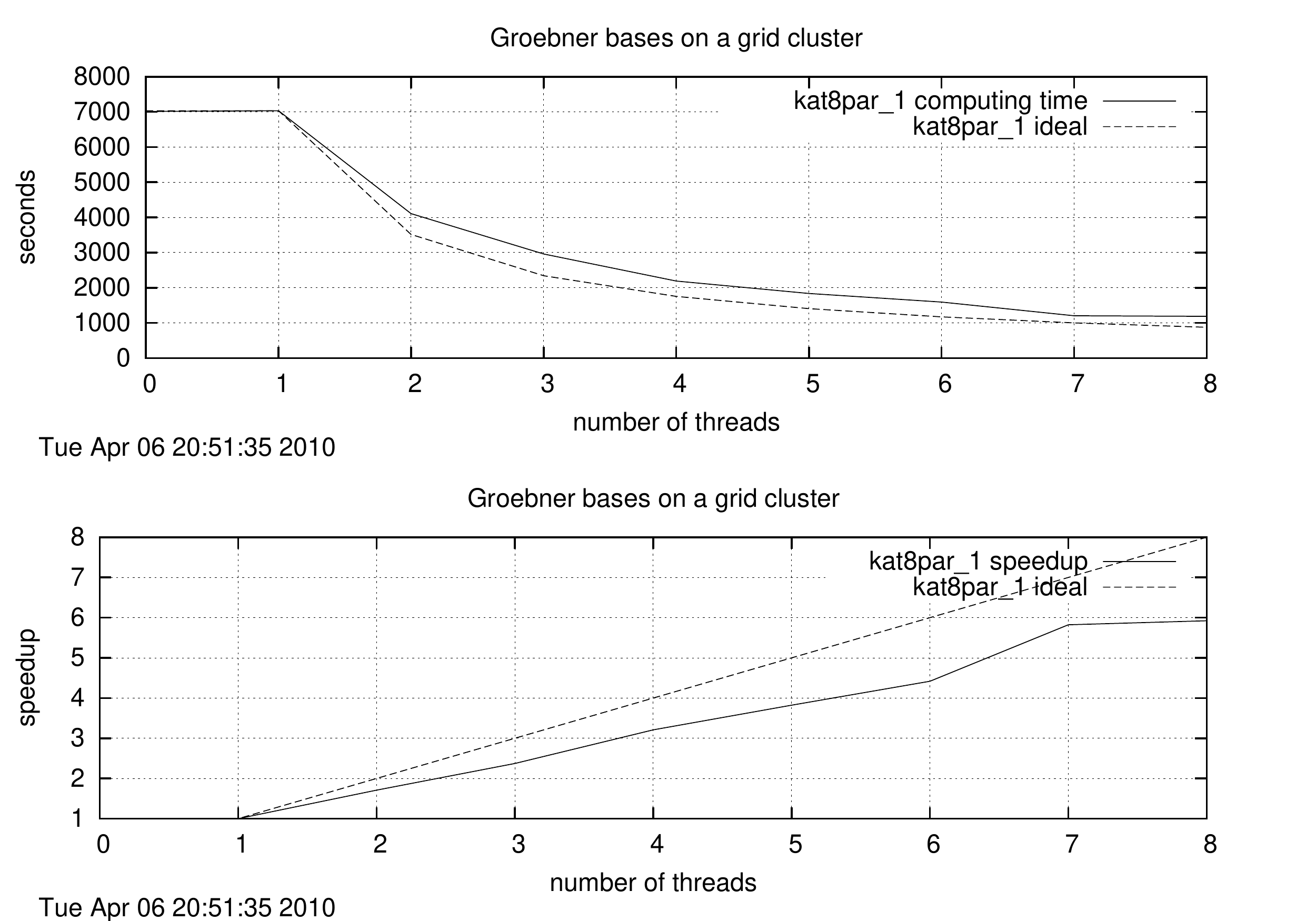,clip=,width=0.70\linewidth,angle=-90}
\caption{Parallel performance, example Katsura 8.}
\label{fig:kat8par_1}
\end{figure}

\begin{table}[thb]
\centering
\caption{Parallel timings for figure \ref{fig:kat8par_1}.}
\label{fig:kat8_par1}
\small{
\include{kat8par_1}
}
\end{table}

\begin{figure}[thb]
\centering
\epsfig{file=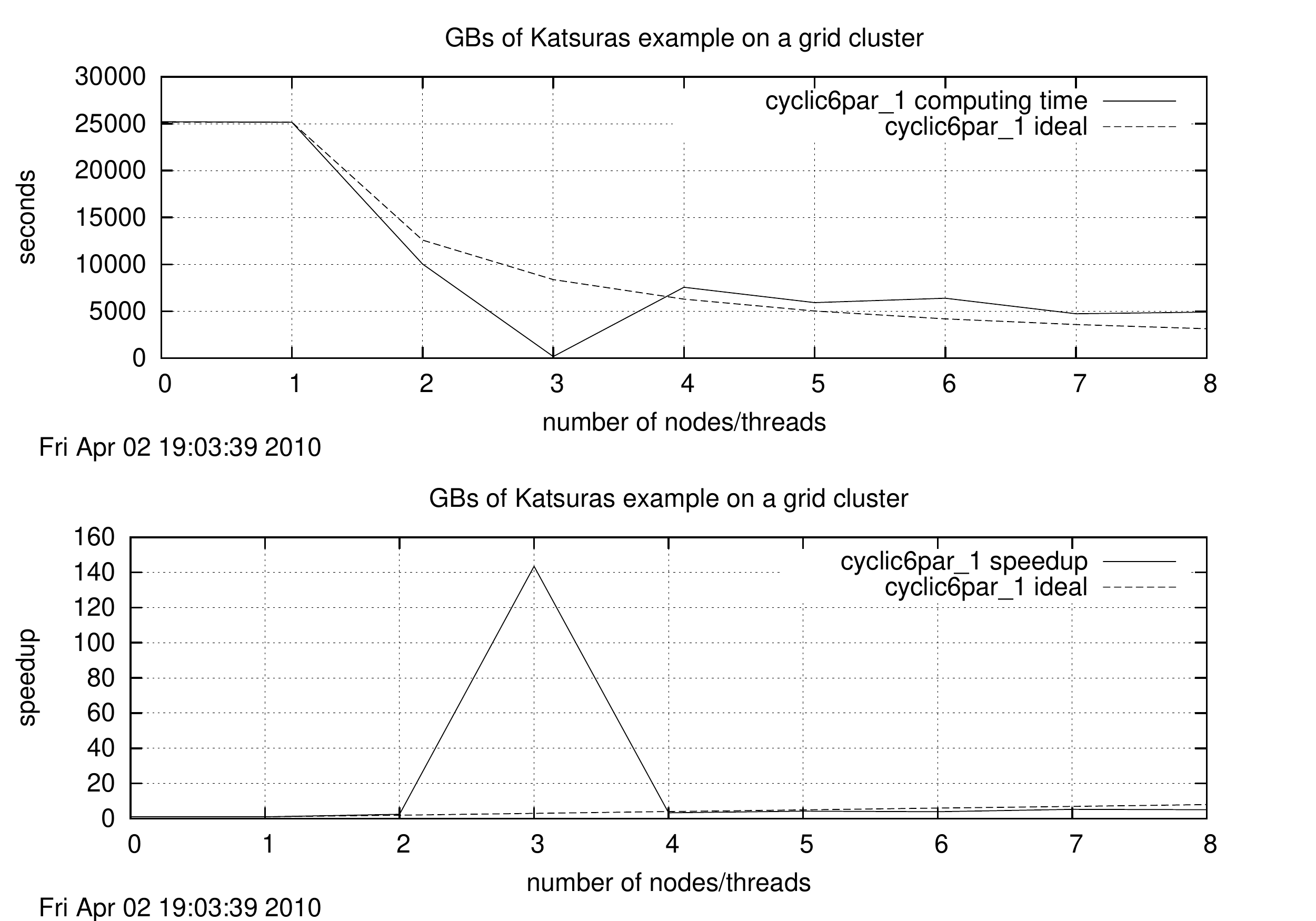,clip=,width=0.70\linewidth,angle=-90}
\caption{Parallel performance, example Cyclic 6.}
\label{fig:c6par1}
\end{figure}

\begin{table}[thb]
\centering
\caption{Parallel timings for figure \ref{fig:c6par1}.}
\label{fig:c61par1t}
\small{
\include{cyclic6par_1}
}
\end{table}

\subsection{Parallel solvable Gr\"obner bases} 

The parallel algorithms are also implemented for solvable polynomial
rings with left, right and two-sided variants.  As in the commutative
case the reductions are performed in parallel by as many threads as
are specified. The right sided Gr\"obner base computation is done via 
the opposite ring and delegation to the left sided parallel computation.
This is not discussed in this paper.

\begin{figure}[thb]
\centering
\epsfig{file=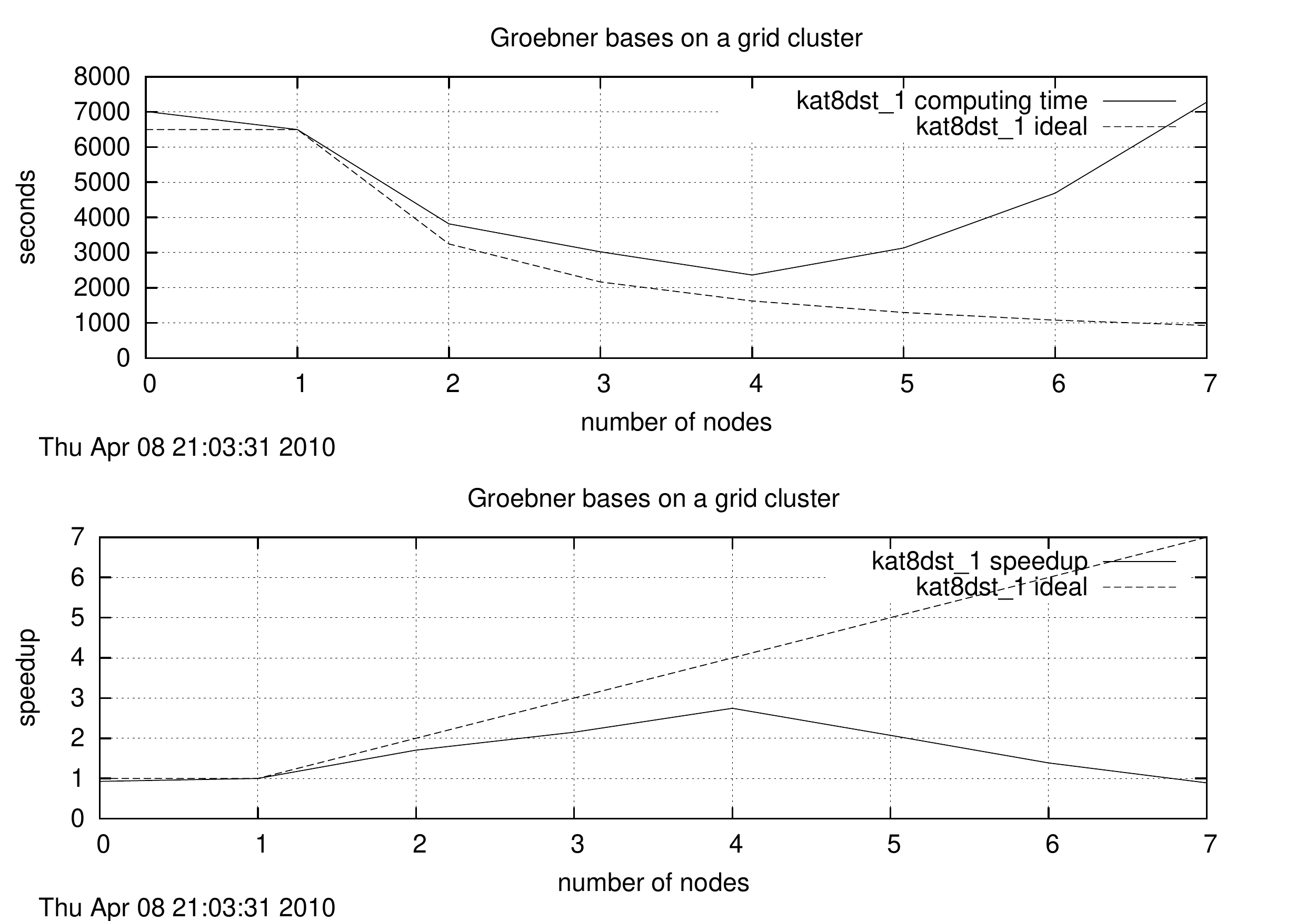,clip=,width=0.70\linewidth,angle=-90}
\caption{Distributed performance, example Katsura 8.}
\label{fig:k8dst1}
\end{figure}

\begin{table}[thb]
\centering
\caption{Distributed timings for figure \ref{fig:k8dst1}.}
\label{fig:k8dst1t}
\small{
\include{kat8dst_1}
}
\end{table}

\begin{figure}[thb]
\centering
\epsfig{file=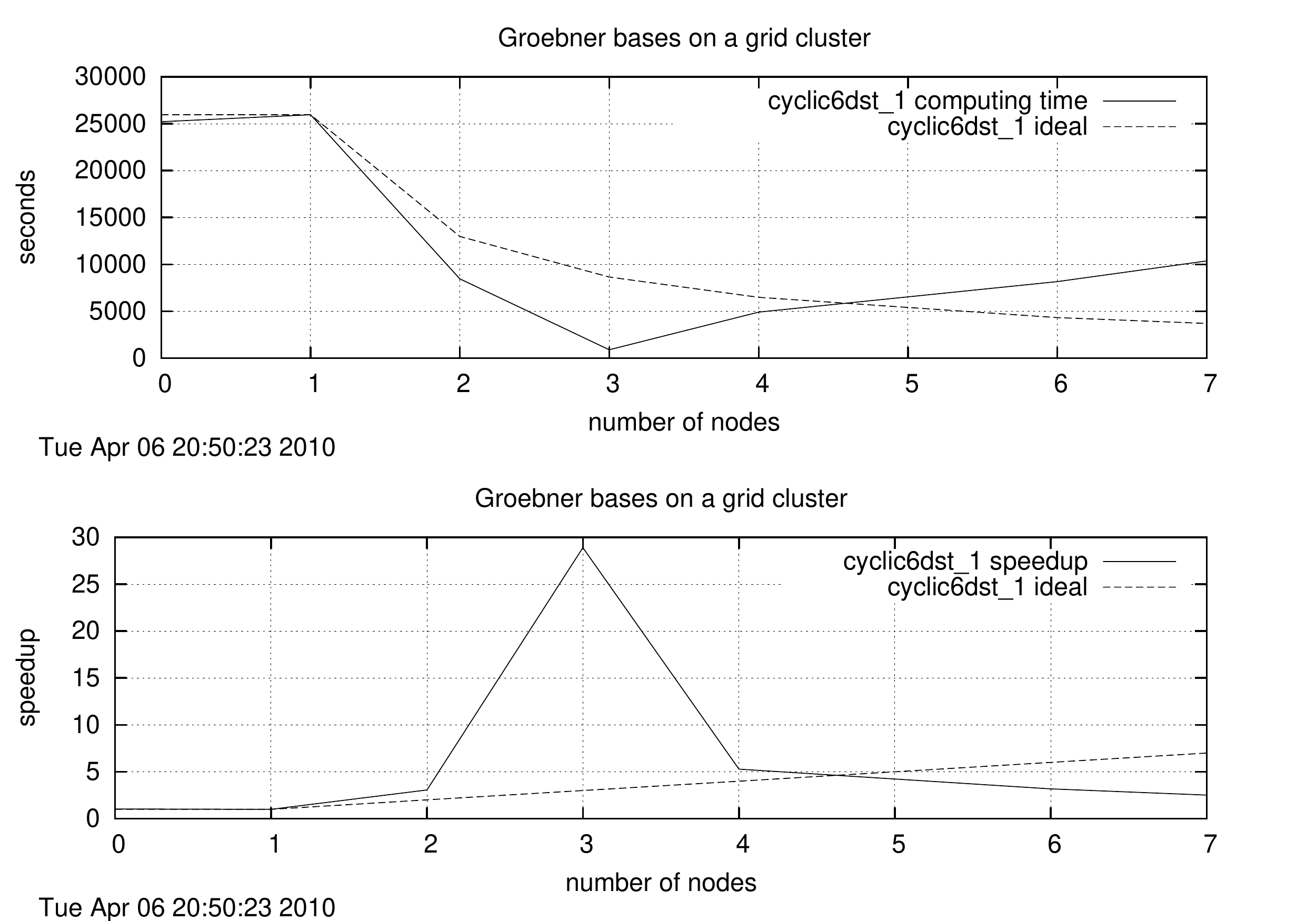,clip=,width=0.70\linewidth,angle=-90}
\caption{Distributed performance, example Cyclic 6.}
\label{fig:c6dst1}
\end{figure}

\begin{table}[thb]
\centering
\caption{Distributed timings for figure \ref{fig:c6dst1}.}
\label{fig:c6dst1t}
\small{
\include{cyclic6dst_1}
}
\end{table}

\subsection{Distributed Gr\"ob\-ner bases} 

We start with the description of the distributed parallel Gr\"obner
base algorithm, the infrastructure required to run it has been
discussed in previous sections.  The description summarizes parts of
\cite{Kredel:2009,Kredel:2010}.  Figure \ref{fig:middleware} gives an
overview of the involved classes and the middle-ware.

The main part of the distributed Gr\"obner bases computation uses the
same work queue (\code{Critical\-Pair\-List} or
\code{Ordered\-Pair\-List}) as the parallel version. From the main
driver method \code{GB()}, shared memory threads take pairs from the
work queue and update the critical pair list by reduced S-polynomials.
But now the threads send the pairs (indexes of pairs) to the
distributed partners for reduction over a network connection and
receive the reduced S-polynomials from the network connection.
Standard Java object serialization is used to encode polynomials for
network transport.

The main method \code{GB()} initializes the critical pair list and the
polynomial list.  The polynomial list is added to a distributed list.
The list index of a polynomial is used as a key in the hash table. The
main method then starts the reducer server threads and waits until
they are terminated.  The reducer servers access the critical pair
list and send pair indexes to the remote reducer client daemons.
Received polynomials are recorded, the critical pair list is updated
and termination conditions are checked.  The reducer client daemons
receive index pairs, performs the reduction and sends the resulting
polynomial back.  Note, only an index of the polynomial in the
distributed list is send, not the polynomial itself.  Only the
reduction result is sent back once to the master and then send to the
distributed lists.  The list is cached on all partner processes and
the master process maintains the polynomials and the index numbers.
The reduction is performed by the distributed processes with the class
\code{Reduction\-Par} which will detect asynchronous updates of the
cached list and restart the reduction from the beginning in such a
case.  

To make use of multiple CPU cores on a node one could start multiple
Java virtual machines (JVM) on it. This approach will however limit
the available memory per JVM, need more TCP/IP connections and will
have higher transport communication overhead.  Also the
\code{Executable\-Server} infrastructure is capable to run multiple
remote jobs in one JVM. This avoids multiple JVMs, but the other
drawbacks remain the same. A better solution is presented in the next
section.

\begin{figure}[thb]
\centering
\epsfig{file=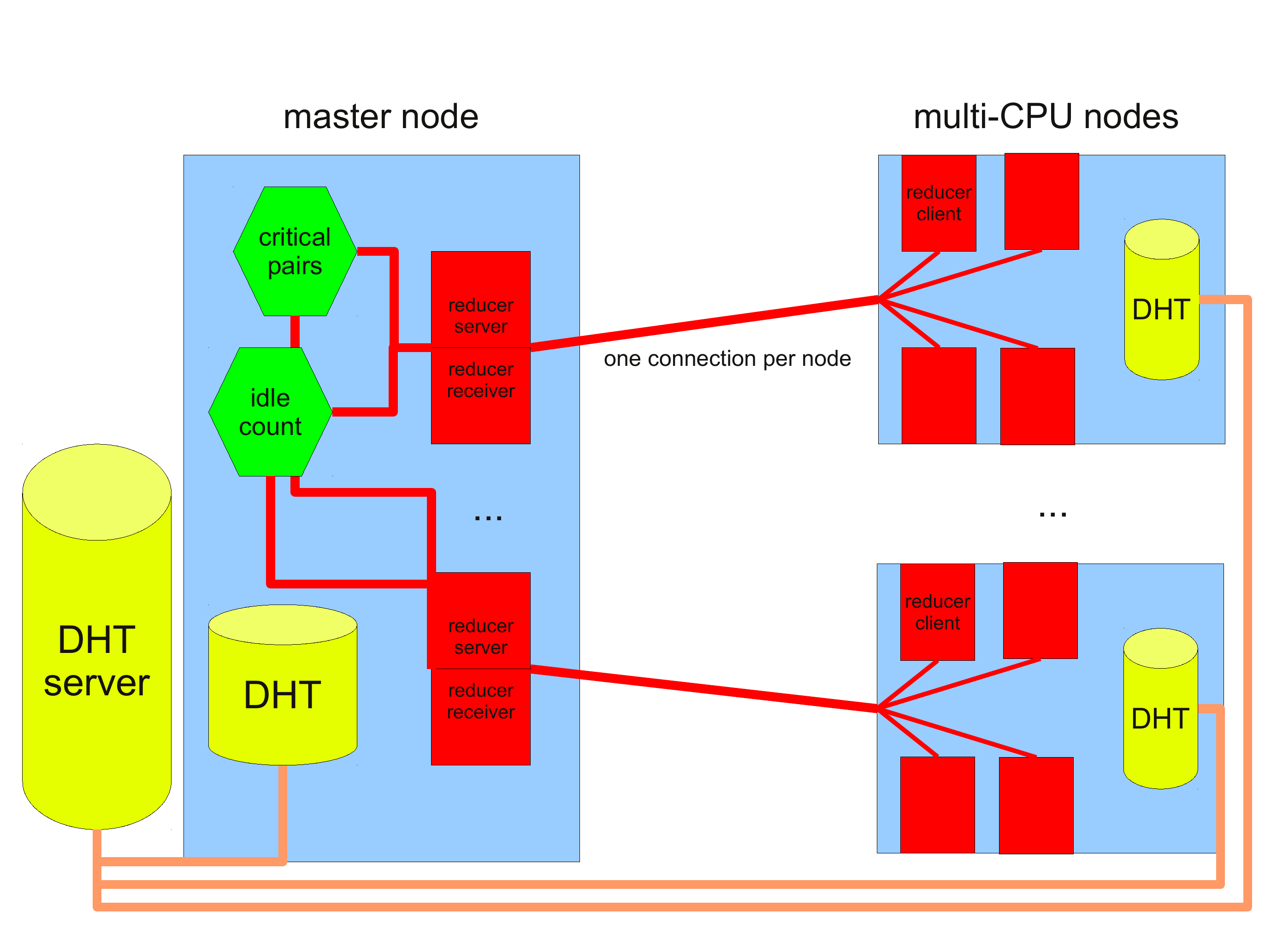,clip=,width=0.95\linewidth}
\caption{Thread to node mapping.}
\label{fig:threadnode}
\end{figure}

\subsection{Distributed hybrid Gr\"ob\-ner ba\-ses} 

In the pure distributed algorithm there is one server thread per
client process on a compute node. In the new hybrid algorithm we have
multiple client threads on a compute node. Looking at figure
\ref{fig:threadnode}, this means, that for the new algorithm multiple
reducer client threads are created.  But there is still only one
reducer server thread per compute node.  The communication for the
pure algorithm is simple: a client requests a pair, the server sends
a pair, the client does the reduction and sends the result back, then
it requests a new pair. Since we now have multiple clients per
communication channel this simple protocol can not be used further. On
the reducer client side we have to extend the protocol: request a
pair, reduce it, send the result back, additionally receive an
acknowledgment then continue to request a new pair.  On the server
side, however, the messages will appear in arbitrary order: pair
request messages will be interleaved with result messages.  To
distinguish between both types of messages we augment messages with
tags representing the respective type. The handling of these tagged
messages is implemented in class \code{Tagged\-Socket\-Channel}.  The
serialized objects send through those channels are tagged with an
unique identifier, which can then be used to receive only certain
messages. The request type messages are received in the main method of
reducer server threads and the result type messages are received
independently in a new separate reducer receiver thread.  So for any
compute node only one communication connection with the master is used
by all threads on the node.

\begin{figure}[thb]
\centering
\epsfig{file=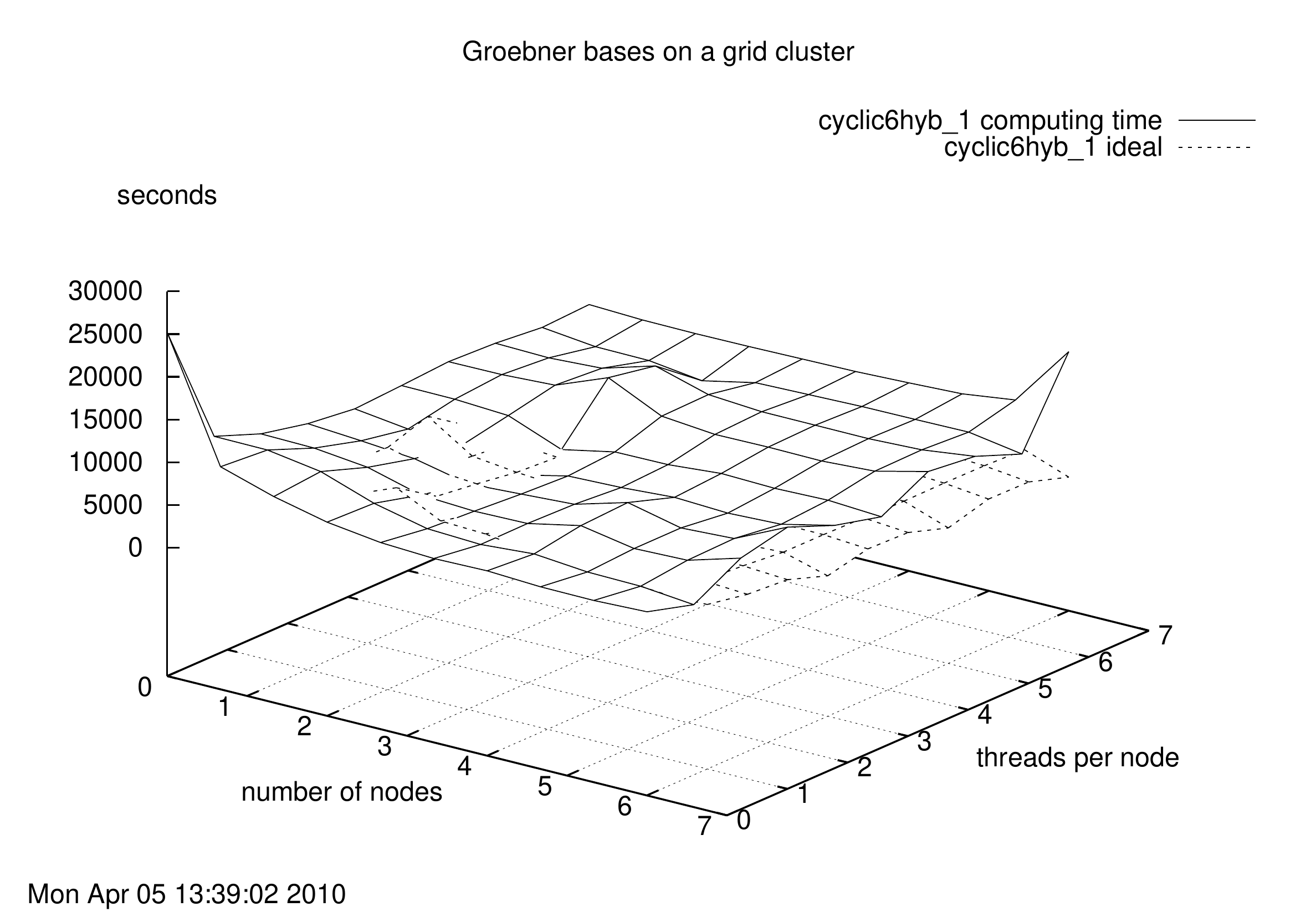,clip=,width=0.70\linewidth,angle=-90}
\caption{Distibuted hybrid performance, example Cyclic 6.}
\label{fig:c61hybrid}
\end{figure}

\begin{table}[thb]
\centering
\caption{Distibuted hybrid timings for figure \ref{fig:c61hybrid}.}
\label{fig:c61_dist1}
\small{
\include{cyclic6hyb_1}
}
\end{table}

\begin{figure}[thb]
\centering
\epsfig{file=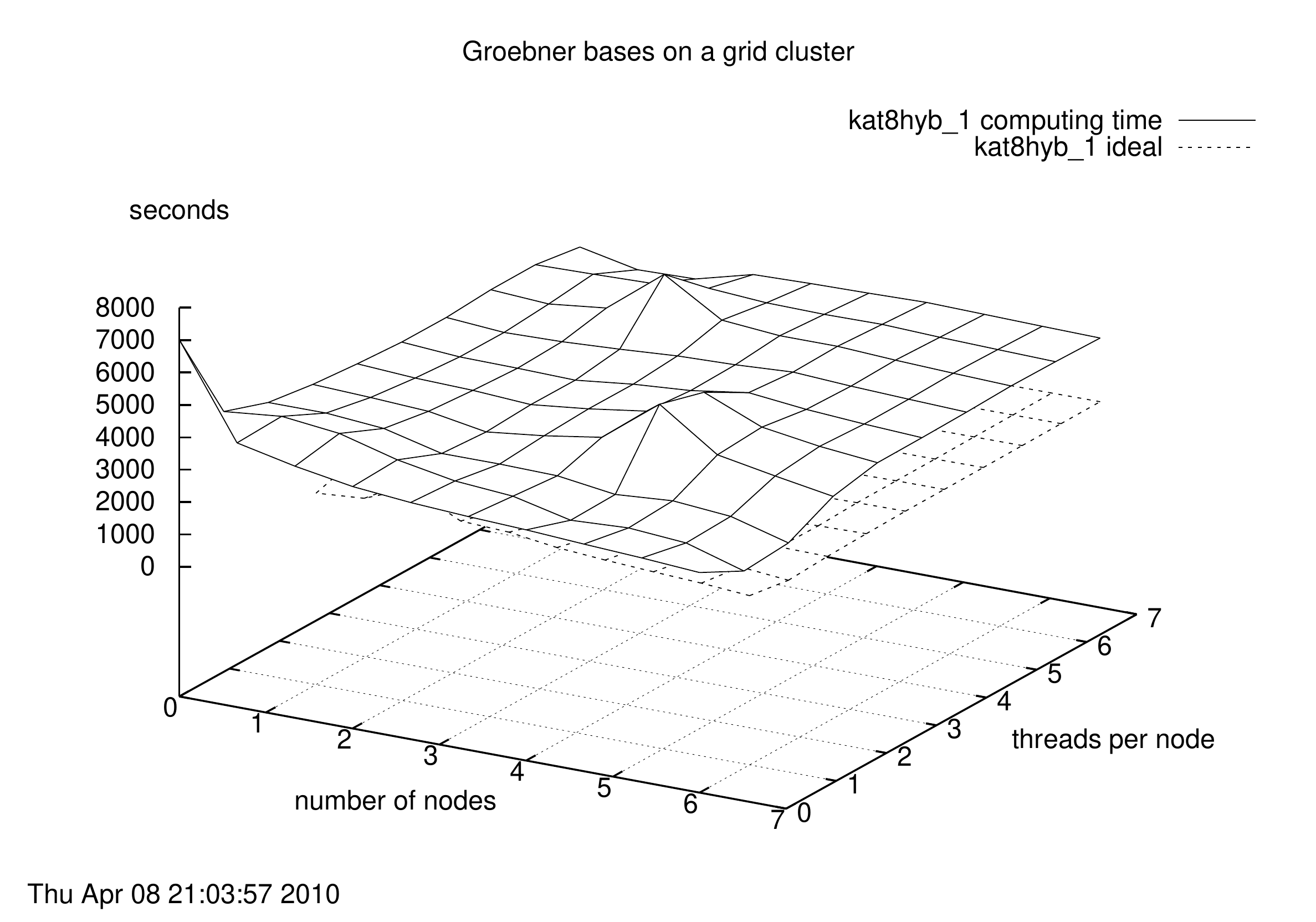,clip=,width=0.70\linewidth,angle=-90}
\caption{Distibuted hybrid performance, example Katsura 8.}
\label{fig:k8hyb1}
\end{figure}

\begin{table}[thb]
\centering
\caption{Distibuted hybrid timings for figure \ref{fig:k8hyb1}.}
\label{fig:k8hyb1t}
\small{
\include{kat8hyb_1}
}
\end{table}

\begin{figure}[thb]
\centering
\epsfig{file=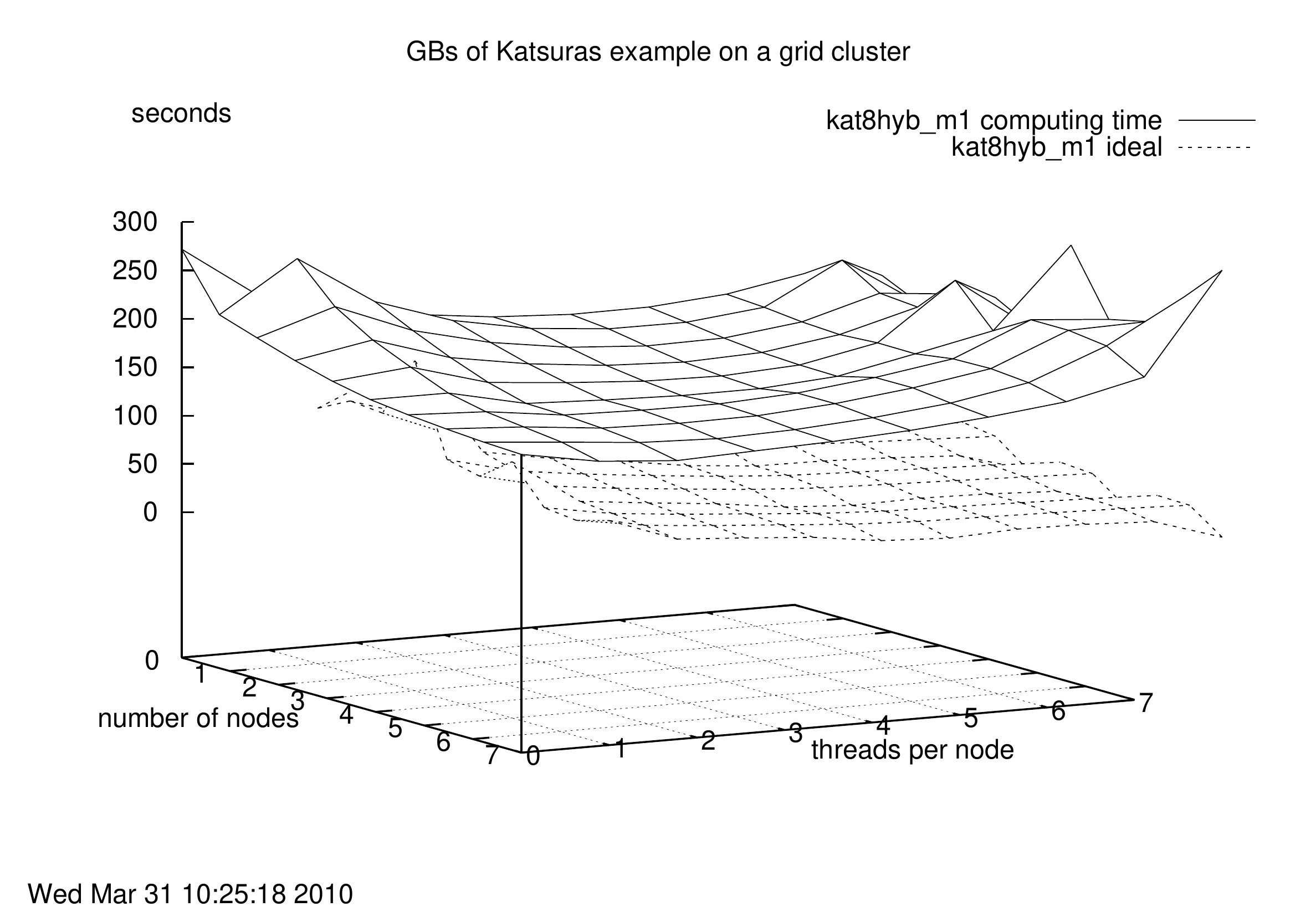,clip=,width=0.70\linewidth,angle=-90}
\caption{Distibuted hybrid performance, example Katsura modulo $2^{127}-1$.}
\label{fig:k8m1hybrid}
\end{figure}

\begin{table}[thb]
\centering
\caption{Distributed hybrid timings for figure \ref{fig:k8m1hybrid}.}
\label{fig:kat8_distm1}
\small{
\include{kat8hyb_m1}
}
\end{table}

\begin{figure}[thb]
\centering
\epsfig{file=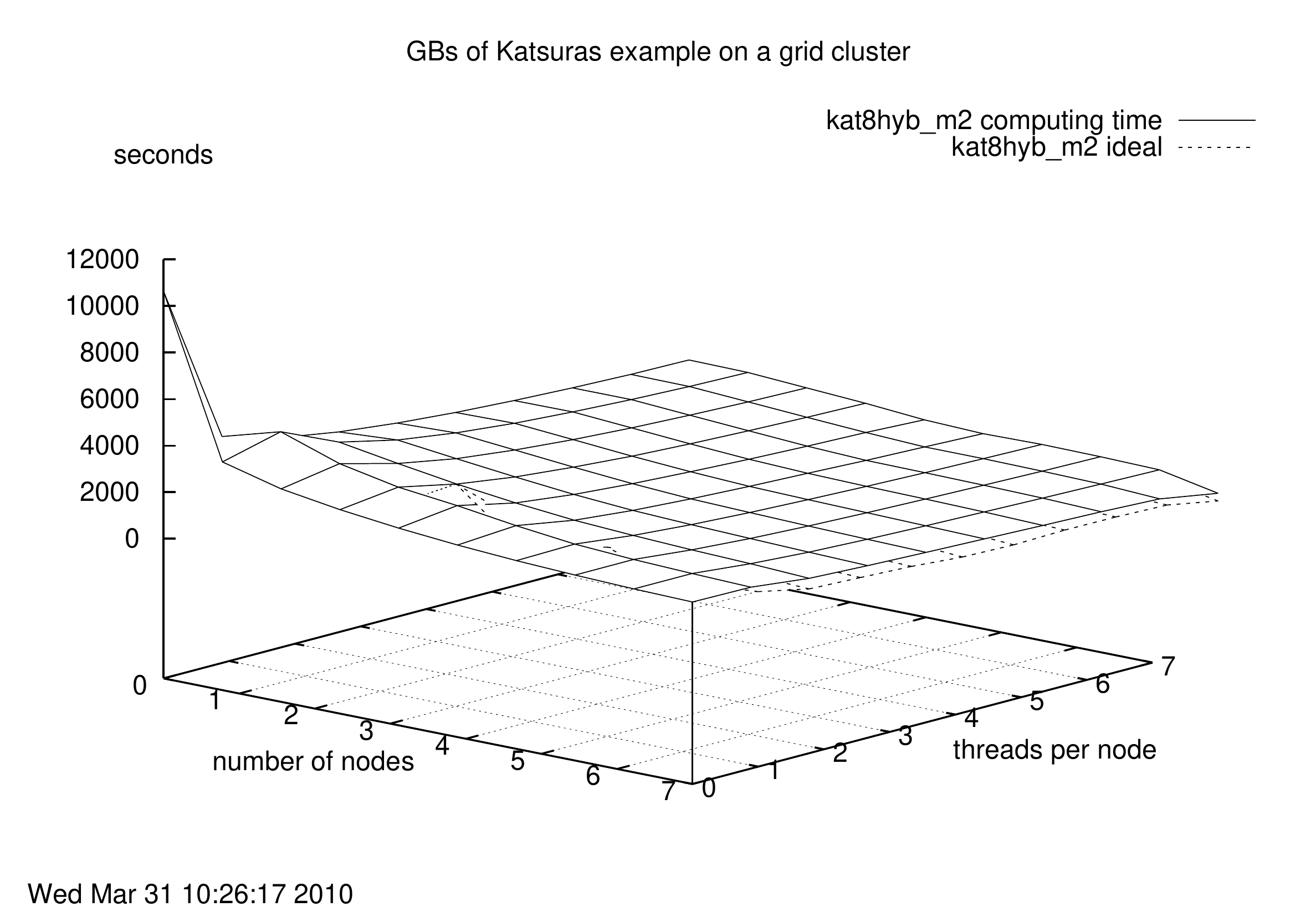,clip=,width=0.70\linewidth,angle=-90}
\caption{Distibuted hybrid performance, example Katsura modulo $2^{3217}-1$.}
\label{fig:k8m2hybrid}
\end{figure}

\begin{table}[thb]
\centering
\caption{Distributed hybrid timings for figure \ref{fig:k8m2hybrid}.}
\label{fig:kat8_distm2}
\small{
\include{kat8hyb_m2}
}
\end{table}

\section{Evaluation} 

In this section we present termination and performance related issues.

\subsection{Termination} 
\label{sec:termi}

In this section we sketch the termination detection in the Buchberger
algorithm. For details see \cite{Kredel:2010}.  As the number of
polynomials in the bases changes and as a consequence the number of
critical pairs changes during the progress of the algorithm, there is
no a-priori way to find out when the algorithm will terminate. Only
the non-constructive proof of Dickson's lemma guarantees, that it will
terminate at all.  Termination is implicitly detected, when all
critical pairs have been processed.

For the sequential algorithm, where we have only one thread of
control, the test if all critical pairs have been processed is
sufficient for termination detection. In a multi-threaded setting this
no longer holds. For example, all but the last thread might find the
set of critical pairs being empty. However, a last thread running
might produce a non-zero reduction polynomial from which a cascade of
new critical pairs could be produced.  So if multiple threads are
used, the termination condition consists of two parts:
\begin{enumerate}
\item the set of critical pairs to process is empty and
\item all threads are idle and not processing a polynomial.
\end{enumerate}
Both conditions have to be checked in a consistent way.

\begin{figure}[thb]
\centering
\epsfig{file=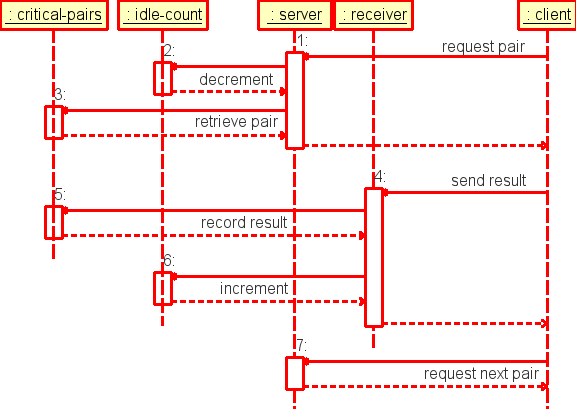,clip=,width=0.95\linewidth}
\caption{Termination of the hybrid GB algorithm.}
\label{fig:hybridtermination}
\end{figure}

The set of critical pairs serves as work queue. They are synchronized
for concurrent access but do not block if they are empty. In case the
set of critical pairs is empty the methods return \code{null}.  In the
hybrid distributed algorithm a thread of the master process is
responsible for more than one distributed thread.  The processing
sequence is shown in figure \ref{fig:hybridtermination}. Condition 2
is ensured by the atomic \code{idle-count}, see also figure
\ref{fig:threadnode}.

\subsection{Performance} 
\label{:sec:perf}

The measurements shown in this paper have all been taken on the
hardware described in section \ref{sec:hardware} and with JDK 1.6.0
with 64-bit server JVM, running with $9$ -- $13$ GB memory, and with
JAS version 2.3, revision 2988.  The examples can be found in
\cite{Graebe:2006}.  The figures show the computing time in seconds
for a given number of threads or nodes, or nodes with threads per node.
In the 2-d plots we show also the speedup.  The corresponding tables
show the number of nodes, the threads/processes per node (ppn), the
computing time in milli-seconds and the speedup. The last two columns
show the number of polynomials put to the critical pair list (put) and
the number of pairs removed from the critical pair list (rem) after
application of the criteria to avoid unnecessary reductions.  The
timings for the sequential algorithm are included with $0$ nodes and
$0$ threads in the figures and tables. To better study the influence
of the transport overhead, the master node is always separated and not
counted.
%
The coefficient rings in the examples are the rational numbers (using
rational arithmetic in coefficients), or modular numbers, if a modulus
is shown.  As modulus we use Mersenne prime 12, $2^{127}-1$ with 39
digits and Mersenne prime 18, $2^{3217}-1$ with 969 digits.  The case
of rational coefficients using fraction free integer arithmetic with
taking primitive parts of polynomials remains to be studied.

Figures \ref{fig:kat8par_1} and \ref{fig:c6par1} show timings and
speedup for the parallel shared memory version of the algorithms.  We
achieve a speedup of $5$ to $6$ using $6$ or $7$ CPUs. This is quite
reasonable, as we run with $2$ garbage collection threads, which
interfere with the computation when all CPUs are occupied.  Figure
\ref{fig:c6par1} shows a speedup of $143$ for $3$ threads by some luck.
Only $177$ polynomials are added in this case to the intermediate ideal
bases instead of about $300$ in the other runs. One could of course also
experience bad luck and hit particular long intermediate ideal bases.
See also the next section \ref{sec:workpara}.

Timings and speedup for the (pure) distributed algorithm is shown in
figures \ref{fig:k8dst1} and \ref{fig:c6dst1}. 
Figure \ref{fig:c6dst1} shows a well behaving example with some
speedup up to $5$ nodes and an extra speedup for $3$ nodes. This time,
however, the number of intermediate polynomials is high ($318$) but the
number of critical pairs remaining after application of the criteria
is only $750$. The example of figure \ref{fig:k8dst1} shows bad speedup
and an extra speed-down for $7$ nodes.

Figures \ref{fig:c61hybrid}, \ref{fig:k8hyb1}, \ref{fig:k8m1hybrid}
and \ref{fig:k8m2hybrid} show timings for the distributed hybrid
algorithms.  Example Cyclic 6 is shown in figure \ref{fig:c61hybrid}.
We see some extra high speedup of $269$ for $2$ nodes and $6$ threads
per node, and of $50$ for $5$ nodes and $2$ threads per node. However,
there is also a speed-down of $0.85$ for $2$ nodes and $5$ threads per
node.  Example Katsura 8 in figure \ref{fig:k8hyb1} shows speedups of
$7$ for $4$ nodes with $2$ threads per node and $6$ nodes with $2$
threads per node.  A speed-down of $0.57$ is observed for $4$ nodes
with $3$ threads per node.  Example Katsura 8 with modular arithmetic
is shown in figures \ref{fig:k8m1hybrid} and \ref{fig:k8m2hybrid}.
The example with a $969$-digit modulus shows very smooth timings and
predictable reasonable speedup of about $2$0 on $5$ nodes using $40$
CPUs, although the absolute computing times are high.  For the
$39$-digit modulus we only see a speedup of $7$ for $2$ nodes with $5$
threads per node and no particular bad speed-down. For even larger
modulus with $6002$ digits, Mersenne prime $2^{19937}-1$, the smooth
timings are lost and more unpredictable timings return (no figure for
this case).  A closer look at tables \ref{fig:c61_dist1},
\ref{fig:k8hyb1t} and \ref{fig:kat8_distm1} 
shows an overhead between $150$ and $300$ seconds for the distributed
hybrid version compared to the sequential version.

In summary we see that the parallel and distributed hybrid algorithms
perform well. The (pure) distributed algorithm is not particular good.
This indicates, that for optimal performance we need to use as many
shared memory CPUs as feasible. For $8$ CPUs on a node it is fast for
up to $6$ threads. Since we use $2$ garbage collection threads on a
node, this is quite reasonable. The communication overhead in the
distributed hybrid algorithm is quite low, as can be seen from the
differences of less than $5 \%$ between the sequential version and the
distributed hybrid version. This is due to the separate distributed
data structure for polynomials with asynchronous updates which avoids
the transport of polynomials as much as possible. Also the
serialization overhead for transport is minimized by the use of
marshalled objects in the distributed data structure. The scaling
obtained in figure \ref{fig:k8m2hybrid} (and table
\ref{fig:kat8_distm2}) also shows that the implementation, the
middle-ware and the infra-structure are quite well performing and are
a good basis for further optimizations regarding selection strategies
and critical pair reductions.

We have not studied the influence of JIT optimizations in this paper.
Our previous measurements \cite{Kredel:2008a,Kredel:2008b,Kredel:2009a} 
show time improvements to $\frac{1}{2}$ and to $\frac{1}{3}$ for
second and third runs of the same example in the same JVM. In this
paper we used fresh JVMs for each run.
%
%

For a discussion of further influencing factors, such as polynomial
and coefficient sizes we must refer to \cite{Kredel:2009}.  For
different selection strategies see the next sub-section.  It remains
to study the optimized Gr\"obner base algorithms
\cite{GebauerMoeller:1988,GioviniMora:1991,Faugere:1999,Faugere:2002} in parallel and
distributed versions with this respect.  
 For further measurements of other algorithms see
 \cite{Kredel:2008a,Kredel:2009,Kredel:2010a}.

\subsection{Workload paradox} 
\label{sec:workpara}

As we have shown above and in \cite{Kredel:2006}, the shared memory
parallel implementations scales well for up to 8 CPUs for a `regular'
problem but it scales only to 3-4 nodes for the pure distributed
algorithm \cite{Kredel:2009}.  One reason is the so called `workload
paradox'. It describes the paradox situation that the parallel and
distributed algorithm have sometimes more work to do than the
sequential algorithm.

The problem has been discussed in \cite{Kredel:2009} for the pure
distributed algorithm.  In this paper it can be seen in figures
\ref{fig:kat8par_1} and \ref{fig:c6par1} for the shared memory parallel
algorithm, in figures \ref{fig:k8dst1} and \ref{fig:c6dst1} for the
pure distributed algorithm and in figures \ref{fig:c61hybrid},
\ref{fig:k8hyb1} for the hybrid algorithm.  We see that the number of
polynomials to be considered varies from $275$ to $433$ and even to
$564$ in the worst case (column put). In the consequence the number of
polynomials to be reduced varies from $1577$ to $1717$ in the worst
case (column rem). Therefore the speedup achieved with the parallel
and distributed algorithms is limited in unlucky cases.

\begin{figure}[thb]
\centering
\epsfig{file=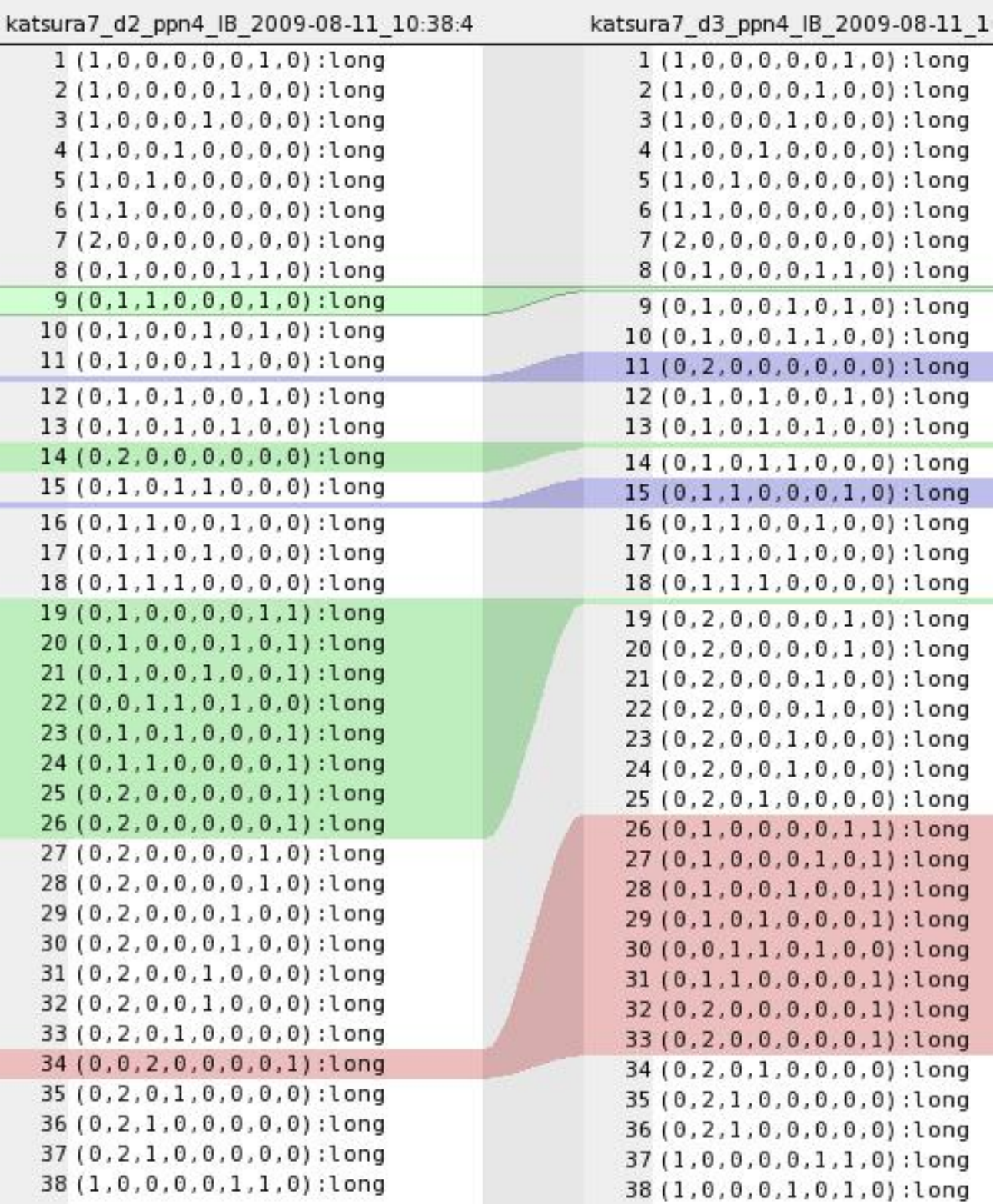,clip=,width=0.5\linewidth}
\mbox{\  }
\epsfig{file=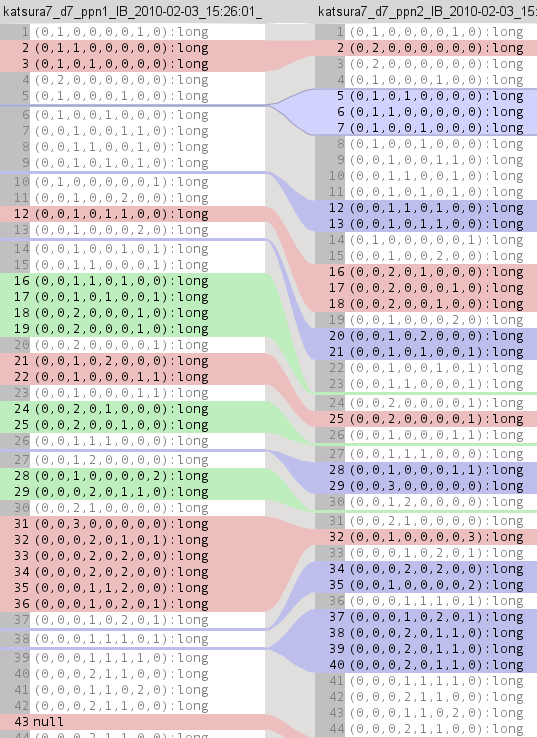,clip=,width=0.45\linewidth}
\caption{Different sequences for pair lcms (left) and reduced polynomials (right).}
\label{fig:lcmpairsdiff}
\label{fig:hpolpairsdiff}
\end{figure}

\subsection{Selection strategies} 
\label{sec:select}

The main work in the Buchberger algorithm is the reduction of a
polynomial with respect to the list of so far computed polynomials.
Already in the sequential algorithm there are several (mathematical)
optimizations to avoid the reduction of polynomials if possible. In
the parallel or distributed algorithm reduced polynomials are found in
a different sequence order since some threads or processes may faster
find polynomials than others. 
See figure \ref{fig:lcmpairsdiff} for an example for two such
sequences represented by the least common multiple of the head terms
of a critical pair and two sequences of head terms of reduced polynomials.
These polynomials are then used to build new pairs for reduction and
so the sequence of polynomials for reduction is most likely different
from the sequential case.

By this observation it seems to be best to use the same order of
polynomials and pairs as in sequential algorithm.  This is to try to
optimize the sequence of critical pairs to be similar to the sequence
in the sequential algorithm \cite{AttardiTraverso:1996}.
However, since the selection algorithm is sequential, any
optimizations eventually reduce the exploitable parallelism and could
also have a negative effect.
In \cite{Amrhein:1996}, the authors discuss two other approaches.  

We have studied two selection strategies.  $n$ reductions are always
performed in parallel.  Then the first `greedy' strategy selects the
first finished result and the second strategy selects the result in
same sequence as reduction has started.  The second strategy is not
yet available for the hybrid algorithm. Although there are examples
where the second strategy is better we found the first strategy to
perform better and to be more robust in other examples.
Due to space limitations we are not able to discuss this topic in more detail,
see the references in section \ref{sec:rela} for an overview of other attempts.

\section{Conclusions} 


We have designed and implemented versions of parallel and distributed
Gr\"obner bases algorithms.  The distributed hybrid algorithm can use
multiple CPUs on the compute nodes and stores the polynomial list only
once per node. There is only one communication channel per node
between the master and the reducer thread on the nodes.  It is usable
and can give considerable speedup for `regular' examples and certain
node numbers and CPUs per node numbers.  The sometimes higher workload
in the parallel and distributed algorithms -- the workload paradox --
limits the applicability in unlucky constellations.  We have also shown
that the implementation, the middle-ware and the infra-structure are
quite well performing and are a good basis for further optimizations.
The implementation fits into the designed hierarchy of Gr\"obner bases
classes and the classes are type-safe designed with Java's generic
types and work for all (implemented exact) fields.

As we have written in \cite{Kredel:2010}, future topics to explore,
include the study of the run-time behavior of the algorithm and
optimized variants, the investigation of different grid middle-wares,
the evaluation of direct Infini\-Band communication, to improve
robustness against node failures or bad reductions. As mentioned in
section \ref{sec:rela} there are many (also mathematical) improvements
and optimizations for the sequential Gr\"obner bases algorithm. These
improvements are hard (eventually impossible) to be carried over to an
parallel algorithm which and is a topic of ongoing research in the area.
A possible parallelization method which has not been studied up to now
is on a higher level.  It is known that the computation of Gr\"obner
bases highly depends on the chosen term ordering.  So a possible
algorithm could start the computation with respect to several term
orderings and use `good' intermediate results from each computation.
The computation of comprehensive Gr\"obner bases
could be parallelized by computing the subtrees on different threads 
\cite{Weispfenning:1992,Kredel:1992}.

\section*{Acknowledgments} 

I thank Thomas Becker for discussions on the implementation of a
polynomial template library and Raphael Jolly for the fruitful
cooperation on the generic type system suitable for a computer algebra
system.  We thank also the referees of \cite{Kredel:2009} and the
participants of the talk for suggestions to improve the paper and the
algorithms.  Thanks also to Markus Aleksy for encouraging this work and
\cite{bwgrid:2008} for providing computing time.

%

\bibliographystyle{abbrv}
\bibliography{jas}


\end{document}

%% file: kat8par_1.tex
\begin{tabular}{|r|r|r|r|r|r|}
\hline
 nodes & ppn & time & speedup & put & rem 
\\ \hline
 0 & 0 & 7013543 & 1.00 & 264 & 1577
\\ \hline
 1 & 1 & 7033558 & 1.0 & 264 & 1577
\\ \hline
 1 & 2 & 4110781 & 1.71 & 266 & 1580
\\ \hline
 1 & 3 & 2962286 & 2.37 & 269 & 1584
\\ \hline
 1 & 4 & 2193242 & 3.20 & 267 & 1579
\\ \hline
 1 & 5 & 1839130 & 3.82 & 279 & 1593
\\ \hline
 1 & 6 & 1592364 & 4.41 & 277 & 1590
\\ \hline
 1 & 7 & 1207742 & 5.82 & 275 & 1586
\\ \hline
 1 & 8 & 1187999 & 5.92 & 272 & 1585
\\ \hline
\end{tabular}

%% file: cyclic6par_1.tex
\begin{tabular}{|r|r|r|r|r|r|}
\hline
\#nodes & ppn & time & speedup & put & rem 
\\ \hline
 0 & 0 & 25200153 & 0.99 & 290 & 1068
\\ \hline
 1 & 1 & 25155926 & 1.0 & 290 & 1068
\\ \hline
 1 & 2 & 10031631 & 2.50 & 296 & 1074
\\ \hline
 1 & 3 & 175335   & 143. & 177 & 606
\\ \hline
 1 & 4 & 7554914  & 3.32 & 310 & 1089
\\ \hline
 1 & 5 & 5916170  & 4.25 & 300 & 1079
\\ \hline
 1 & 6 & 6382993  & 3.94 & 301 & 1079
\\ \hline
 1 & 7 & 4736758  & 5.31 & 309 & 1087
\\ \hline
 1 & 8 & 4907370  & 5.12 & 239 & 826
\\ \hline
\end{tabular}

%% file: kat8dst_1.tex
\begin{tabular}{|r|r|r|r|r|r|}
\hline
 nodes & ppn & time & speedup & put & rem 
\\ \hline
 0 & 0 & 7013543 & 0.92 & 264 & 1577
\\ \hline
 1 & 1 & 6497092 & 1.0 & 267 & 1580
\\ \hline
 2 & 1 & 3816389 & 1.70 & 292 & 1606
\\ \hline
 3 & 1 & 3023442 & 2.14 & 311 & 1628
\\ \hline
 4 & 1 & 2365191 & 2.74 & 339 & 1655
\\ \hline
 5 & 1 & 3131358 & 2.07 & 326 & 1642
\\ \hline
 6 & 1 & 4690347 & 1.38 & 331 & 1648
\\ \hline
 7 & 1 & 7275751 & 0.89 & 374 & 1693
\\ \hline
\end{tabular}

%% file: cyclic6dst_1.tex
\begin{tabular}{|r|r|r|r|r|r|}
\hline
 nodes & ppn & time & speedup & put & rem 
\\ \hline
 0 & 0 & 25200153 & 1.02 & 290 & 1068
\\ \hline
 1 & 1 & 25941962 & 1.0 & 297 & 1075
\\ \hline
 2 & 1 & 8464227 & 3.06 & 305 & 890
\\ \hline
 3 & 1 & 898151 & 28.8 & 318 & 750
\\ \hline
 4 & 1 & 4916018 & 5.27 & 366 & 958
\\ \hline
 6 & 1 & 8161999 & 3.17 & 415 & 1202
\\ \hline
 7 & 1 & 10359709 & 2.50 & 427 & 1213
\\ \hline
\end{tabular}

%% file: cyclic6hyb_1.tex
\begin{tabular}{|r|r|r|r|r|r|}
\hline
 nodes & ppn & time & speedup & put & rem 
\\ \hline
 0 & 0 & 25200153 & 1.00 & 290 & 1068
\\ \hline
 1 & 1 & 25354798 & 1.0 & 290 & 1068
\\ \hline
 1 & 2 & 6870749  & 3.69 & 261 & 848
\\ \hline
 1 & 3 & 945055   & 26.8 & 261 & 689
\\ \hline
 1 & 4 & 9351502  & 2.71 & 385 & 1164
\\ \hline
 1 & 5 & 8119667  & 3.12 & 433 & 1216
\\ \hline
 1 & 6 & 689379   & 36.7 & 321 & 758
\\ \hline
 2 & 1 & 8289801  & 3.05 & 283 & 868
\\ \hline
 2 & 2 & 222789   & 113. & 295 & 732
\\ \hline
 2 & 3 & 478916   & 52.9 & 324 & 757
\\ \hline
 2 & 4 & 522992   & 48.4 & 339 & 785
\\ \hline
 2 & 5 & 29531917 & 0.85 & 411 & 1194
\\ \hline
 2 & 6 & 93922    & 269. & 384 & 819
\\ \hline
 3 & 1 & 187588   & 135. & 275 & 703
\\ \hline
 3 & 2 & 161671   & 156. & 342 & 774
\\ \hline
 3 & 3 & 2127834  & 11.9 & 304 & 685
\\ \hline
 3 & 4 & 2411978  & 10.5 & 374 & 841
\\ \hline
 4 & 1 & 5833243  & 4.34 & 335 & 920
\\ \hline
 4 & 2 & 15810128 & 1.60 & 325 & 763
\\ \hline
 4 & 3 & 5117109  & 4.95 & 410 & 1191
\\ \hline
 5 & 1 & 1324398  & 19.1 & 303 & 734
\\ \hline
 5 & 2 & 498929   & 50.8 & 305 & 746
\\ \hline
 5 & 3 & 510808   & 49.6 & 358 & 806
\\ \hline
 5 & 4 & 4661828  & 5.43 & 397 & 924
\\ \hline
 6 & 1 & 374217   & 67.7 & 292 & 725
\\ \hline
 6 & 2 & 21742099 & 1.16 & 335 & 927
\\ \hline
 6 & 3 & 7275150  & 3.48 & 441 & 997
\\ \hline
 6 & 4 & 3758169  & 6.74 & 439 & 1219
\\ \hline
 6 & 5 & 15974842 & 1.58 & 497 & 1281
\\ \hline
 6 & 6 & 647052   & 39.1 & 564 & 1364
\\ \hline
 6 & 7 & 15300911 & 1.65 & 563 & 1239
\\ \hline
\end{tabular}

%% file: kat8hyb_1.tex
\begin{tabular}{|r|r|r|r|r|r|}
\hline
 nodes & ppn & time & speedup & put & rem 
\\ \hline
 0 & 0 & 7013543 & 1.04 & 264 & 1577
\\ \hline
 1 & 1 & 7317628 & 1.0 & 264 & 1577
\\ \hline
 1 & 2 & 3964386 & 1.84 & 283 & 1597
\\ \hline
 1 & 3 & 2875307 & 2.54 & 296 & 1609
\\ \hline
 1 & 4 & 2291950 & 3.19 & 317 & 1630
\\ \hline
 1 & 5 & 1969253 & 3.71 & 332 & 1646
\\ \hline
 1 & 6 & 1812295 & 4.03 & 347 & 1662
\\ \hline
 1 & 7 & 3965623 & 1.84 & 331 & 1641
\\ \hline
 1 & 8 & 1422994 & 5.14 & 347 & 1662
\\ \hline
 2 & 1 & 3724833 & 1.96 & 285 & 1598
\\ \hline
 2 & 2 & 2276379 & 3.21 & 311 & 1628
\\ \hline
 2 & 3 & 1536457 & 4.76 & 346 & 1660
\\ \hline
 2 & 4 & 2636121 & 2.77 & 343 & 1661
\\ \hline
 2 & 5 & 1243440 & 5.88 & 368 & 1682
\\ \hline
 2 & 6 & 7990697 & 0.91 & 338 & 1664
\\ \hline
 3 & 1 & 3039801 & 2.40 & 309 & 1626
\\ \hline
 3 & 2 & 1708402 & 4.28 & 329 & 1643
\\ \hline
 3 & 3 & 1930677 & 3.79 & 363 & 1679
\\ \hline
 3 & 4 & 3166579 & 2.31 & 359 & 1668
\\ \hline
 3 & 5 & 3439587 & 2.12 & 357 & 1669
\\ \hline
 4 & 1 & 2163145 & 3.38 & 318 & 1634
\\ \hline
 4 & 2 & 1058099 & 6.91 & 348 & 1665
\\ \hline
 4 & 3 & 12648595 & 0.57 & 365 & 1680
\\ \hline
 5 & 1 & 1789862 & 4.08 & 328 & 1641
\\ \hline
 5 & 2 & 1881553 & 3.88 & 357 & 1677
\\ \hline
 5 & 3 & 3465200 & 2.11 & 399 & 1717
\\ \hline
 6 & 1 & 1494186 & 4.89 & 345 & 1662
\\ \hline
 6 & 2 & 1015457 & 7.20 & 382 & 1701
\\ \hline
\end{tabular}

%% file: kat8hyb_m1.tex
\begin{tabular}{|r|r|r|r|r|r|}
\hline
\#nodes & ppn & time & speedup & put & rem 
\\ \hline
 0 & 0 & 271947 & 2.10 & 264 & 1577
\\ \hline
 1 & 1 & 573481 & 1.0 & 264 & 1577
\\ \hline
 1 & 2 & 307770 & 1.86 & 286 & 1600
\\ \hline
 1 & 3 & 202947 & 2.82 & 319 & 1633
\\ \hline
 1 & 4 & 146953 & 3.90 & 330 & 1647
\\ \hline
 1 & 5 & 134112 & 4.27 & 360 & 1675
\\ \hline
 1 & 6 & 108636 & 5.27 & 368 & 1683
\\ \hline
 1 & 7 & 413031 & 1.38 & 353 & 1668
\\ \hline
 1 & 8 & 79480  & 7.21 & 424 & 1735
\\ \hline
 2 & 1 & 321407 & 1.78 & 288 & 1601
\\ \hline
 2 & 2 & 171407 & 3.34 & 328 & 1644
\\ \hline
 2 & 3 & 132093 & 4.34 & 366 & 1683
\\ \hline
 2 & 4 & 96068  & 5.96 & 391 & 1709
\\ \hline
 2 & 5 & 82720  & 6.93 & 420 & 1734
\\ \hline
 2 & 6 & 115711 & 4.95 & 473 & 1787
\\ \hline
 2 & 7 & 127011 & 4.51 & 480 & 1802
\\ \hline
 2 & 8 & 245371 & 2.33 & 374 & 1692
\\ \hline
 3 & 1 & 216913 & 2.64 & 312 & 1629
\\ \hline
 3 & 2 & 128662 & 4.45 & 366 & 1684
\\ \hline
 3 & 3 & 133768 & 4.28 & 401 & 1720
\\ \hline
 3 & 4 & 104162 & 5.50 & 433 & 1752
\\ \hline
 3 & 5 & 90056  & 6.36 & 489 & 1799
\\ \hline
 3 & 6 & 115549 & 4.96 & 559 & 1874
\\ \hline
 3 & 7 & 430092 & 1.33 & 428 & 1741
\\ \hline
 3 & 8 & 114203 & 5.02 & 611 & 1942
\\ \hline
 4 & 1 & 176818 & 3.24 & 328 & 1647
\\ \hline
 4 & 2 & 97581  & 5.87 & 401 & 1718
\\ \hline
 4 & 3 & 110804 & 5.17 & 430 & 1750
\\ \hline
 4 & 4 & 93711  & 6.11 & 479 & 1802
\\ \hline
 4 & 5 & 68071  & 8.42 & 510 & 1834
\\ \hline
 4 & 6 & 149877 & 3.82 & 528 & 1847
\\ \hline
 4 & 7 & 126118 & 4.54 & 556 & 1873
\\ \hline
 4 & 8 & 414340 & 1.38 & 403 & 1722
\\ \hline
 5 & 1 & 147274 & 3.89 & 359 & 1673
\\ \hline
 5 & 2 & 102234 & 5.60 & 414 & 1724
\\ \hline
 5 & 3 & 138203 & 4.14 & 474 & 1792
\\ \hline
 5 & 4 & 105733 & 5.42 & 521 & 1830
\\ \hline
 5 & 5 & 217189 & 2.64 & 457 & 1774
\\ \hline
 5 & 6 & 133299 & 4.30 & 660 & 1986
\\ \hline
 5 & 7 & 427047 & 1.34 & 462 & 1781
\\ \hline
 5 & 8 & 137789 & 4.16 & 696 & 2022
\\ \hline
 6 & 1 & 137391 & 4.17 & 380 & 1695
\\ \hline
 6 & 2 & 107357 & 5.34 & 420 & 1737
\\ \hline
 6 & 3 & 71713  & 7.99 & 519 & 1835
\\ \hline
 6 & 4 & 109077 & 5.25 & 586 & 1912
\\ \hline
 6 & 5 & 107350 & 5.34 & 633 & 1962
\\ \hline
 6 & 6 & 138562 & 4.13 & 867 & 2200
\\ \hline
 6 & 7 & 264173 & 2.17 & 607 & 1916
\\ \hline
 6 & 8 & 421697 & 1.35 & 555 & 1887
\\ \hline
 7 & 1 & 128016 & 4.47 & 394 & 1711
\\ \hline
 7 & 2 & 94762  & 6.05 & 437 & 1757
\\ \hline
 7 & 3 & 110094 & 5.20 & 526 & 1837
\\ \hline
 7 & 4 & 112712 & 5.08 & 637 & 1953
\\ \hline
 7 & 5 & 126195 & 4.54 & 665 & 1985
\\ \hline
 7 & 6 & 132914 & 4.31 & 686 & 2018
\\ \hline
 7 & 7 & 168587 & 3.40 & 1069 & 2425
\\ \hline
 7 & 8 & 286053 & 2.00 & 627 & 1957
\\ \hline
\end{tabular}

%% file: kat8hyb_m2.tex
\begin{tabular}{|r|r|r|r|r|r|}
\hline
\#nodes & ppn & time & speedup & put & rem 
\\ \hline
 0 & 0 & 10634093 & 1.00 & 264 & 1577
\\ \hline
 1 & 1 & 10638363 & 1.0 & 264 & 1577
\\ \hline
 1 & 2 & 5557931  & 1.91 & 280 & 1594
\\ \hline
 1 & 3 & 3901336  & 2.72 & 302 & 1617
\\ \hline
 1 & 4 & 3111197  & 3.41 & 308 & 1624
\\ \hline
 1 & 5 & 2544589  & 4.18 & 323 & 1639
\\ \hline
 1 & 6 & 2191707  & 4.85 & 318 & 1633
\\ \hline
 1 & 7 & 1885106  & 5.64 & 329 & 1645
\\ \hline
 1 & 8 & 1732946  & 6.13 & 333 & 1649
\\ \hline
 2 & 1 & 5614045  & 1.89 & 278 & 1591
\\ \hline
 2 & 2 & 2986583  & 3.56 & 299 & 1615
\\ \hline
 2 & 3 & 2080451  & 5.11 & 327 & 1640
\\ \hline
 2 & 4 & 1632211  & 6.51 & 352 & 1666
\\ \hline
 2 & 5 & 1372935  & 7.74 & 337 & 1654
\\ \hline
 2 & 6 & 1275509  & 8.34 & 421 & 1733
\\ \hline
 2 & 7 & 1027207  & 10.3 & 441 & 1749
\\ \hline
 2 & 8 & 993523  & 10.7 & 484 & 1797
\\ \hline
 3 & 1 & 3922971 & 2.71 & 299 & 1614
\\ \hline
 3 & 2 & 2071164 & 5.13 & 326 & 1641
\\ \hline
 3 & 3 & 1492716 & 7.12 & 335 & 1649
\\ \hline
 3 & 4 & 1169490 & 9.09 & 378 & 1699
\\ \hline
 3 & 5 & 1062725 & 10.0 & 447 & 1756
\\ \hline
 3 & 6 & 929664  & 11.4 & 441 & 1755
\\ \hline
 3 & 7 & 961792  & 11.0 & 484 & 1792
\\ \hline
 3 & 8 & 836431  & 12.7 & 554 & 1869
\\ \hline
 4 & 1 & 3074408 & 3.46 & 312 & 1629
\\ \hline
 4 & 2 & 1634847 & 6.50 & 359 & 1675
\\ \hline
 4 & 3 & 1216382 & 8.74 & 382 & 1695
\\ \hline
 4 & 4 & 978586  & 10.8 & 409 & 1725
\\ \hline
 4 & 5 & 900061  & 11.8 & 511 & 1824
\\ \hline
 4 & 6 & 729377  & 14.5 & 417 & 1719
\\ \hline
 4 & 7 & 687423  & 15.4 & 565 & 1882
\\ \hline
 4 & 8 & 718741  & 14.8 & 576 & 1902
\\ \hline
 5 & 1 & 2475993 & 4.29 & 324 & 1641
\\ \hline
 5 & 2 & 1342804 & 7.92 & 390 & 1707
\\ \hline
 5 & 3 & 1018719 & 10.4 & 386 & 1705
\\ \hline
 5 & 4 & 901174  & 11.8 & 460 & 1779
\\ \hline
 5 & 5 & 792400  & 13.4 & 486 & 1801
\\ \hline
 5 & 6 & 683510  & 15.5 & 543 & 1862
\\ \hline
 5 & 7 & 639776  & 16.6 & 685 & 2009
\\ \hline
 5 & 8 & 507932  & 20.9 & 590 & 1906
\\ \hline
 6 & 1 & 2119696 & 5.01 & 330 & 1644
\\ \hline
 6 & 2 & 1182626 & 8.99 & 393 & 1709
\\ \hline
 6 & 3 & 832588  & 12.7 & 387 & 1698
\\ \hline
 6 & 4 & 773272  & 13.7 & 525 & 1838
\\ \hline
 6 & 5 & 793126  & 13.4 & 470 & 1791
\\ \hline
 6 & 6 & 677274  & 15.7 & 543 & 1851
\\ \hline
 6 & 7 & 578902  & 18.3 & 511 & 1835
\\ \hline
 6 & 8 & 584389  & 18.2 & 849 & 2177
\\ \hline
 7 & 1 & 1817419 & 5.85 & 354 & 1668
\\ \hline
 7 & 2 & 1048401 & 10.1 & 365 & 1678
\\ \hline
 7 & 3 & 849506  & 12.5 & 482 & 1793
\\ \hline
 7 & 4 & 733319  & 14.5 & 607 & 1923
\\ \hline
 7 & 5 & 588179  & 18.0 & 489 & 1805
\\ \hline
 7 & 6 & 580260  & 18.3 & 579 & 1906
\\ \hline
 7 & 7 & 594621  & 17.8 & 723 & 2047
\\ \hline
 7 & 8 & 524730  & 20.2 & 682 & 2006
\\ \hline
\end{tabular}